\documentclass[preprint,times,12pt]{elsarticle}

\usepackage{amsmath}
\usepackage{units}
\usepackage{graphicx}

\newcommand{\etal}{\emph{et al.}}
\newcommand{\ie}{\emph{i.e.}}
\newcommand{\eg}{\emph{e.g.}}

\newcommand{\rff}[1]{\ref{fig:#1}}
\newcommand{\rfe}[1]{Eq.\ref{equ:#1}}
\newcommand{\ud}{\mathrm{d}}
\newcommand{\deriv}[2]{\frac{\ud #1}{\ud #2}}
\newcommand{\pderiv}[2]{\frac{\partial #1}{\partial #2}}
\newcommand{\ppderiv}[3]{\frac{\partial^2 #1}{\partial #2 \partial #3}}

\newcommand{\apar}{\ensuremath{\alpha}}

\begin{document}

\begin{frontmatter}
  
  \title{The global covariance matrix of tracks fitted with a Kalman
    filter and an application in detector alignment}


\author{W.~D.~Hulsbergen}
\ead{wouter.hulsberg@nikhef.nl}
\address{National Institute for Subatomic Physics (Nikhef), Amsterdam, The Netherlands.}

\begin{abstract}
  We present an expression for the covariance matrix of the set of
  state vectors describing a track fitted with a Kalman filter. We
  demonstrate that this expression facilitates the use of a Kalman
  filter track model in a minimum $\chi^2$ algorithm for the alignment
  of tracking detectors. We also show that it allows to incorporate
  vertex constraints in such a procedure without refitting the tracks.
\end{abstract}

\end{frontmatter}


\section{Introduction}

Minimum $\chi^2$ algorithms for the alignment of tracking detectors
generally come in two flavours, namely those that ignore and those
that do not ignore the correlations between hit residuals. The former
are sometimes called \emph{local} or \emph{iterative} methods while
the latter are called \emph{global} or \emph{closed-form}
methods~\cite{Blobel:2006yh}.  The advantage of the closed-form
methods is that for an alignment problem in which the measurement
model is a linear function of both track and alignment parameters the
solution that minimizes the total $\chi^2$ can be obtained with a
single pass over the data.

The covariance matrix for the track parameters is an essential
ingredient to the closed-form alignment approach~\cite{Blobel:2002ax}.
If the track fit is performed using the standard expression for the
least-squares estimator (sometimes called the \emph{standard} or
\emph{global} fit method), the computation of the covariance matrix is
a natural part of the track fit. This is why previously reported
implementations of the closed-form alignment procedure
(e.g.~\cite{Belotelov:2006su, Kleinwort:2006zz, Bruckman:2005,
  Bocci:2007zzb,Flucke:2008zz, Viret:2008jq}) make use of the standard
fit.

In contrast most modern particle physics experiments rely on a Kalman
filter track fit~\cite{Kalman:1960,Fruhwirth:1987fm} for default track
reconstruction. The Kalman filter is less computationally expensive
than the standard fit and facilitates an easy treatment of multiple
scattering in the form of process noise. However, the computation of
the covariance matrix in the common Kalman track fit is not complete:
The correlations between track parameters at different position along
the track are not calculated. In the presence of process noise these
correlations are non-trivial. Consequently, the result of the common
Kalman track fit cannot be used directly in a closed-form alignment
procedure.

In this paper we present the expressions for the computation of the
global covariance matrix --- the covariance matrix for all parameters
in the track model --- in a Kalman filter track fit. We show how this
result can be used in an alignment procedure. Furthermore, using
similar expressions we demonstrate how vertex constraints can be
applied in the alignment without refitting the tracks in the vertex.
To illustrate that our approach leads to a functional closed-form
alignment algorithm, we present some results obtained for the
alignment of the LHCb vertex detector with Monte Carlo simulated data.

An important motivation for extending the Kalman track fit for use in
a closed-form alignment approach is that the estimation of alignment
parameters is not independent of the track model. Typically, in
closed-form alignment procedures the track model used in the alignment
is different from that used in the track reconstruction for physics
analysis, which in practise is always a Kalman filter.  Sometimes the
track model in the alignment is simplified, ignoring multiple
scattering corrections or the magnetic field. The imperfections in the
track model used for alignment will partially be absorbed in
calibration parameters.  Consequently, in order the guarantee
consistency between track model and detector alignment, it is
desirable to use the default track fit in the alignment procedure.

The Kalman filter has also been proposed for the estimation of the
alignment parameters themselves~\cite{Widl:2007zza}. This method for
alignment is an alternative formulation of the closed-form alignment
approach that is particularly attractive if the number of alignment
parameter is large.  Our results for the global covariance matrix of
the Kalman filter track model and for vertex constraints can
eventually be applied in such a Kalman filter alignment procedure.


\section{Minimum $\chi^2$ formalism for alignment}

To show that the global covariance matrix of the track parameters is
an essential ingredient to the closed-form alignment approach, we
briefly revisit the minimum $\chi^2$ formalism for alignment. Consider
a track $\chi^2$ defined as
\begin{equation}
  \chi^2 \; = \; \left[ \rule{0ex}{1.7ex} m - h(x) \right]^T \: V^{-1} \: \left[ \rule{0ex}{1.7ex} m - h(x) \right] ,
  \label{equ:trackchisquare}
\end{equation}
where $m$ is a vector of measured coordinates, $V$ is a (usually
diagonal) covariance matrix, $h(x)$ is the measurement model and $x$
is the vector of track parameters. Note that~\rfe{trackchisquare} is a
matrix expression: $m$ and $h$ are vectors and $V$ is a symmetric
matrix, all with dimension equal to the number of measurements.

For a linear expansion of the measurement model around an initial
estimate $x_0$ of the track parameters,
\[
h(x) = h(x_0) + H (x - x_0),
\]
where
\[
    H \; = \; \left.\pderiv{h(x)}{x}\right|_{x_0}
\]
is sometimes called the derivative or projection matrix, the condition
that the $\chi^2$ be minimal with respect to $x$ can be written as
\[
    0 \; \equiv \; \deriv{\chi^2}{x} \; = \; - 2 H^T V^{-1} \left[ m - h(x_0) - H ( x-x_0) \rule{0ex}{1.7ex} \right] .
    \label{equ:minchisq}
\]
The solution to this system of equations is given by the well known
expression for the least squares estimator
\begin{equation}
  x \; = \; x_0  -  C H^{T} V^{-1} \left[ m - h(x_0) \rule{0ex}{1.7ex} \right] ,
  \label{equ:lsm}
\end{equation}
where the matrix $C$ is the covariance matrix for $x$
\begin{equation}
  C \; = \; \left( H^{T} V^{-1} H \right)^{-1} .
  \label{equ:lsmcov}
\end{equation}
If the measurement model is not linear, \ie{} if $H$ depends on $x$,
expression~\rfe{lsm} can be applied iteratively, until a certain
convergence criterion is met, for example defined by a minimum change
in the $\chi^2$. In that case it makes sense to write~\rfe{lsm} in
terms of the first and second derivative of the $\chi^2$ at the
current estimate $x_0$
\begin{equation*}
  x - x_0 \; = - \left( \left.\deriv{^2\chi^2}{x^2}\right|_{x_0} \right)^{-1} \left.\deriv{\chi^2}{x}\right|_{x_0}
  \label{equ:newtonraphson}
\end{equation*}
and regard the iterative minimization procedure as an application of
the Newton-Raphson method.

We now consider an extension of the measurement model with a set of
calibration parameters $\apar$,
\[
   h(x) \rightarrow h(x,\apar) .
\]
The parameters $\apar$ are considered common to all tracks in a
particular calibration sample. We estimate $\apar$ by minimizing the
sum of the $\chi^2$ values of the tracks simultaneously with respect
to $\apar$ and the track parameters $x_i$ of each track $i$,
\begin{equation}
  \pderiv{\sum_i \chi_i^2}{\apar} \; = \; 0 \qquad \text{and} \qquad 
  \forall_{i} \; \pderiv{\chi_i^2}{x_i} \; = \; 0 .
\end{equation}
Please, note that the index $i$ refers to the track and not to a
component of the vector $x$. We will omit the index from now on and
consider only the $\chi^2$ contribution from a single track.

The number of parameters in the minimization problem above scales with
the number of tracks. If the number of tracks is large enough, a
computation that uses an expression for the least squares estimator
analogous to~\rfe{lsm} is computationally too expensive. A more
practical method relies on a computation in two steps. First, track
parameters are estimated for an initial set of calibration parameters
$\apar_0$.  Subsequently, the total $\chi^2$ is minimized with respect
to $\apar$ taking into account the dependence of $x_i$ on $\apar$,
\eg{} through the total derivative
\begin{equation}
   \deriv{}{\apar} \; = \; \pderiv{}{\apar} \: + \: \deriv{x}{\apar} \pderiv{}{x} .
   \label{equ:ddalphaA}
\end{equation}

The derivative matrix $\ud x/\ud\apar$ in~\rfe{ddalphaA} follows from
the condition that the $\chi^2$ of the track remains minimal with
respect to $x$, which can be expressed as
\[
   \deriv{}{\apar} \pderiv{\chi^2}{x} \; = \; 0
\]
and results in
\begin{equation}
  \deriv{x}{\apar} \; = \; - \ppderiv{\chi^2}{\apar}{x} \left( \pderiv{^2\chi^2}{x^2} \right)^{-1} .
  \label{equ:dxdalpha}
\end{equation}
Note that if the problem is linear this derivative is independent of
the actual value of $x$ or $\apar$. Consequently, in this limit this
expression remains valid even if the track $\chi^2$ was not yet
minimized with respect to $x$.

The condition that the total $\chi^2$ of a sample of tracks be minimal
with respect to both track and alignment parameters can now be
expressed as
\begin{equation}
  0 \; \equiv \; \deriv{\chi^2}{\apar} 
  \label{equ:minchisqconditionforalignment}
\end{equation}
For $M$ alignment parameter this defines a system of $M$ coupled
non-linear equations. In analogy with the procedure introduced for the
track $\chi^2$ minimization above we search for a solution by
linearizing the minimum $\chi^2$ condition around an initial value
$\apar_0$ and solving the linear system of $M$ equations
\begin{equation}
    \left.\deriv{^2\chi^2}{\apar^2}\right|_{\apar_0} \Delta\apar \; 
    = \; - \left.\deriv{\chi^2}{\apar}\right|_{\apar_0}
    \label{equ:globalequation}
\end{equation}
for $\Delta\apar$. In the remainder of this section we derive the
expressions for these derivatives.

To simplify the notation we define the residual vector of the track
\[
  r \;=\; m - h(x,\apar)
\]
and its derivative to $\apar$
\[
    A_{k\ell} \; \equiv \; \pderiv{r_k}{\apar_\ell} .
\]
We linearize $r$ around the expansion point $(x(\apar_0),\apar_0)$,
and using~\rfe{dxdalpha} obtain for any total derivative to $\apar$
\[
  \deriv{}{\apar} \; = \; \pderiv{}{\apar} - A^{T} V^{-1} H C \pderiv{}{x} .
  \label{equ:ddalphaB}
\]
(The minus sign appears because $H$ is the derivative of $h$ and not
of $r$.)  In this expression we have substituted the covariance matrix
for $C$ for $x$.  The first and second derivatives of the $\chi^2$
contribution of a single track are now given by
\begin{equation}
  \deriv{\chi^2}{\apar} \; = \; 2 A^{T} V^{-1} \left( V - H C H^{T} \right) V^{-1} r ,
  \label{equ:firstdchisqda}
\end{equation}
\begin{equation}
  \deriv{^2\chi^2}{\apar^2} \; = \; 2  A^{T}  V^{-1} \left( V -  H C H^{T} \right) V^{-1} A .
  \label{equ:seconddchisqda}
\end{equation}
The matrix
\begin{equation}
   R \; \equiv \; V - HCH^T
   \label{equ:residualcovariance}
\end{equation}
that appears in these expressions is the covariance matrix for the
residuals $r$. This matrix is in general singular and its rank is the
number of degrees of freedom of the fit.

If the track parameters $x$ for which the residuals $r$ and $H$ are
calculated, are actually those that minimize the track's $\chi^2$ for
the current set of alignment constants $\apar_0$, the residuals satisfy
the least squares condition $H^T V^{-1} r = 0 $ and the first
derivative to $\apar$ reduces to
\begin{equation}
   \deriv{\chi^2}{\apar} \; = \; 2 A^{T} V^{-1} r .
   \label{equ:firstdchisqdaminchisq}
\end{equation}
Consequently, if $V$ is diagonal, the derivative to a particular
parameter $\apar_j$ only receives contributions from residuals for
which $\partial r_i/ \partial \apar_j$ does not vanish.\footnote{In
  other words, if $\apar_i$ is an alignment parameter of module $X$,
  only hits in module $X$ contribute to the first derivative of the
  $\chi^2$ to $\apar_i$.}  An important consequence of this is that if
there are additional contributions to the tracks $\chi^2$, in
particular hits in subdetectors that we do not align for, constraints
from a vertex fit or multiple scattering terms, then these terms only
enter the derivative calculation through the track covariance matrix
$C$. We will exploit this property in the next section when we discuss
the use of a Kalman filter track model for alignment.

The expressions~\rfe{firstdchisqdaminchisq} and~\rfe{seconddchisqda}
can now be used to evaluate the first and second derivative for an
initial calibration $\apar_0$ over a given track sample and inserted
in~\rfe{globalequation} to obtain an improved calibration. If the
residuals are non-linear in either track parameters or alignment
parameters, several iterations may be necessary to minimize the
$\chi^2$.

If the alignment is sufficiently constrained, the second derivative
matrix can be inverted and the covariance matrix for the alignment
parameters is given by
\[
  \text{Cov}(\apar) \; = \; 2 \left( \deriv{^2\chi^2}{\apar^2} \right)^{-1}.
\]
Ignoring higher order derivatives in $\apar$, the change in the total
$\chi^2$ as the result of a change $\Delta\apar$ in the alignment
parameters can be written as
\[
\Delta\chi^2 \; = \; \frac{1}{2} \: \deriv{\chi^2}{\apar}^T \: \Delta\apar
             \; = \; - \Delta\apar^T \: {\text{Cov}(\apar)}^{-1} \: \Delta\apar .
\]
Consequently, the change in the total $\chi^2$ is equivalent to the
significance of the alignment correction. The quantity $\Delta\chi^2$
is a useful measure for following the convergence of an alignment.


\section{The global covariance matrix in the Kalman filter track fit \label{sec:kalmancorrelation}}

In the global method for track fitting a track is modelled by a single
$N$ parameter vector (usually $N=5$) at a fixed position along the
track. Multiple scattering can be incorporated in this model by
introducing explicit parameters for the kinks at scattering planes.
The parameters that minimize the $\chi^2$ and the corresponding
covariance matrix follow from the application of the least squares
estimator~\rfe{lsm}.

In the Kalman filter method for track fitting~\cite{Fruhwirth:1987fm}
a track is modelled by a separate $N$ dimensional track parameter
vector (or \emph{state vector}) at each measurement (or \emph{node}).
The state vectors are related by a \emph{transport} function, which
follows from the equation of motion of the charged particle.  In the
absence of multiple scattering the state vectors are one-to-one
functions of one-another and hence fully correlated. In the presence
of multiple scattering the correlation is reduced by introducing
so-called process noise in the propagation of the state vector between
neighbouring nodes.

As we have seen in the previous section the closed-form method for
alignment uses the vector of residuals $r$ and a corresponding
covariance matrix $R$. The covariance matrix for the residuals can be
computed from the global covariance matrix of the track parameters.
However, the correlations between the state vectors at different nodes
are normally not calculated in the Kalman filter: They are either not
computed at all (if the smoothing is done as a weighted average of a
forward and backward filter) or (if the Rauch-Tung-Striebel smoother
formalism is applied) only implicitly and only between neighbouring
nodes.

To derive an expression for the covariance matrix of all parameters in
the Kalman filter track model we use the notation of
reference~\cite{Fruhwirth:1987fm} for the linear Kalman filter, in
particular
\begin{itemize}
\item $x_{k}$ is the state vector at node $k$ after accumulating the
  information from measurements $\{1, \ldots, k\}$;
\item $C_{k}$ is the covariance of $x_{k}$;
\item $x_{k}^n$ is the state vector at node $k$ after processing all
  $n$ measurements.
\end{itemize}
In the following we first calculate the correlation between $x_{k-1}$
and $x_{k}$, which we denote by $C_{k-1,k}$. From this we proceed with
the correlation matrix between $x_{k-1}^{n}$ and $x_{k}^{n}$.  The
correlation between any two states $k$ and $l$ then follows from the
observation that the correlation between these states occurs via
intermediate states.

In the notation of~\cite{Fruhwirth:1987fm} we have for the prediction
of state $k$ from state $k-1$,
\[
  x_k^{k-1} \; = \; F_{k-1} x_{k-1} ,
\]
where $F$ is the Jacobian or transport matrix. The covariance of the
prediction is given by
\[
  C_{k}^{k-1} \; = \; F_{k-1} C_{k-1} F_{k-1}^{T}  \; + \; Q_{k-1}
\]
where $Q_{k-1}$ is the process noise in the transition from state
$k-1$ to $k$. The full covariance matrix for the pair of states
$(x_{k-1}, x_k^{k-1})$ is then given by
\[
  \text{Cov}(x_{k-1}, x_k^{k-1}) \; = \;
  \left( 
    \begin{array}{cc} 
       C_{k-1}             &  C_{k-1}  F_{k-1}^{T}           \\   
      \;F_{k-1} C_{k-1} \;\;  & \; F_{k-1} C_{k-1} F_{k-1}^{T} + Q_{k-1}\; 
    \end{array}
  \right)
\]
In the Kalman filter track fit we now proceed by adding the
information of measurement $k$ to obtain a new estimate for the state
in procedure that is called \emph{filtering} and leads to state vector
$x_k$. The remaining measurements ${k+1,\ldots,n}$ are processed with
prediction and filter steps in the same fashion. Afterwards a
procedure called \emph{smoothing} can be applied to recursively
propagate the information obtained through measurements
${k+1,\ldots,n}$ back to node $k$. The smoothed state vector at node
$k$ is labelled by $x_k^{n}$ and its covariance by $C_{k}^{n}$. To
derive the expression for the covariance matrix of the smoothed states
$x_k^{n}$ and $x_{k-1}^{n}$ we first present the following lemma.

Suppose we have two observables $a$ and $b$ with covariance matrix
\[
     \left(\begin{array}{cc} V_{aa} & V_{ab} \\ V_{ba} & V_{bb} \\ \end{array} \right)
\]
Now suppose we have obtained a new estimate of $\tilde{a}$ with
variance $\tilde{V}_{aa}$ by adding information. We can propagate the
new information to $b$ with a least squares estimator, which gives
\begin{equation}
  \begin{array}{ccl}
    \tilde{b} & = & b + V_{ba} V_{aa}^{-1} ( \tilde{a} - a ) \\
    \tilde{V}_{bb} & = & V_{bb} + V_{ba} V_{aa}^{-1} ( \tilde{V}_{aa} - V_{aa} ) V_{aa}^{-1} V_{ab} \\
    \tilde{V}_{ab} & = & V_{ab} + ( \tilde{V}_{aa} - V_{aa} ) V_{aa}^{-1} V_{ab} \; = \; \tilde{V}_{aa} V_{aa}^{-1} V_{ab}
  \end{array}
  \label{equ:propagation}
\end{equation}
This expression also holds if $a$ and $b$ are vectors. It can be
derived by minimizing the following $\chi^2$
\newcommand{\twovec}[2]{\left( \begin{array}{c} #1 \\ #2 \end{array}
  \right)}
\[
  \chi^2 \; = \; 
   \twovec{\tilde{a} - a }{\tilde{b} - b }^{T} 
   \left( \begin{array}{cc} V_{aa} & V_{ab} \\ V_{ba} & V_{bb} \\ \end{array} \right)^{-1}
   \twovec{\tilde{a} - a }{\tilde{b} - b} + \left(\tilde{a} - m\right)^T {V_m}^{-1} \left(\tilde{a} - m\right)
\]
(where $m$ with variance $V_m$ is the additional information for $a$)
with respect to $\tilde{a}$ and $\tilde{b}$.

Substituting $x_{k-1}$ for $b$, $x_k$ for $a$ and $x_k^{n}$ for
$\tilde{a}$ in~\rfe{propagation} we obtain for the correlation between
the smoothed states
\begin{equation}
  C_{k-1,k}^{n} \; = \; C_{k-1}  F_{k-1}^{T} \left( C_{k}^{k-1} \right)^{-1} C_{k}^{n} \; = \; A_{k-1} C_{k}^{n}
  \label{equ:correlation}
\end{equation}
where we have used the definition of the smoother gain
matrix~\cite{Fruhwirth:1987fm}
\begin{equation}
  A_{k-1} \; = \; C_{k-1} F_{k-1}^T \left( C_{k}^{k-1} \right)^{-1} .
  \label{equ:smoothergain}
\end{equation}
For the smoothed state $x_{k-1}^{n}$ and its covariance we find
\[
\begin{array}{ccl}
  x_{k-1}^{n} & = & x_{k-1} + A_{k-1} ( x_k^n - x_k^{k-1} ), \\
  C_{k-1}^{n} & = &  C_{k-1} \; + \; A_{k-1} \left( C_{k}^{n} - C_{k}^{k-1} \right) A_{k-1}^{T}.
\end{array}
\]
These are the Rauch-Tung-Striebel smoothing expressions as found
in~\cite{Fruhwirth:1987fm}. The gain matrix in~\rfe{smoothergain} can
be written in different forms, \eg{}
\[
A_{k-1} \; = \; \left( F_{k-1} \right)^{-1} \left(C_{k}^{k-1} - Q_{k-1} \right)\left(C_{k}^{k-1}\right)^{-1} .
\]
This expression shows explicitly that $A_{k-1} = \left( F_{k-1}
\right)^{-1}$ if there is no process noise ($Q=0$). Therefore, as one
expects, without process noise the smoothed states in the Kalman
filter are just related by the transport equation.

Once we have the calculated the off-diagonal element $(k-1,k)$, we
proceed to the next diagonal $(k-2,k)$. The correlation between states
$k-2$ and $k$ can be calculated by performing a simultaneous smoothing
of states $k-2$ and $k-1$: In the argument above we substitute a new
vector $(x_{k-2},x_{k-1})$ for $b$, rather than just the state
$x_{k-1}$. The result for the correlation $(k-2,k)$ is
\[
  C_{k-2,k}^{n} \; = \; C_{k-2,k-1}^{n} \left( C_{k-1}^{n} \right)^{-1} C_{k-1,k}^{n} .
  \label{equ:correlationkmintwok}
\]
This expression can also be derived with a simpler argument: The
origin of the correlation between state vectors is the transport.
Therefore, the correlation $(k-2,k)$ occurs only through the
correlations $(k-2,k-1)$ and $(k-1,k)$. Following the same reasoning
the next diagonal becomes
\[
\begin{array}{ccl}
C_{k-3,k}^{n} 
 & = & C_{k-3,k-2}^{n} \left( C_{k-2}^{n} \right)^{-1} C_{k-2,k-1}^{n} \left( C_{k-1}^{n} \right)^{-1} C_{k-1,k}^{n} \\
 & = & C_{k-3,k-2}^{n} \left( C_{k-2}^{n} \right)^{-1} C_{k-2,k}^{n} ,
\end{array}
\]
which shows that the calculation can be performed recursively. By
substituting the smoother gain matrix we can write this in the
following compact form
\begin{equation}
C_{k-1,\ell}^{n} 
\; = \; C_{k-1,k}^{n} \left( C_{k}^{n} \right)^{-1} C_{k,\ell}^{n}
\; = \; A_{k-1} C_{k,\ell}^{n} \qquad k \leq \ell .
\label{equ:recursivecorrelation}
\end{equation}
If the gain matrices are temporarily stored, then if $N$ is the
dimension of the state vector, the calculation of each off-diagonal
element in the full covariance matrix of state vectors requires about
$N$ multiplications and $N$ additions.  The total number of operations
by far exceeds the numerical complication of the standard Kalman
filter. However, we have found that for tracks traversing the entire
LHCb tracking system, with a total of about 30 measurement
coordinates, the computational cost for the global covariance matrix
with the procedure above was smaller than that of the Kalman filter
track fit itself. This is because the LHCb track fit is largely
dominated by integration of the inhomogeneous magnetic field and the
location of intersections with detector material.

Now that we have calculated the full covariance matrix of all states
$x_k^{(n)}$, the elements of the covariance matrix $R$ of the
residuals are simply given by
\begin{equation}
  \label{equ:kalmanfullR}
  R_{k\ell} \; = \; V_{k}\delta_{kl} - H_k C_{k,\ell}^n H_\ell^{T} .
\end{equation}
This completes the recipe for using a Kalman filter track model in
the alignment of tracking detectors.

We have argued below~\rfe{firstdchisqdaminchisq} that the cancellation
that takes place between~\rfe{firstdchisqda}
and~\rfe{firstdchisqdaminchisq} is important when considering the
Kalman track fit for alignment.  This can be explained as follows. If
we were to use the Kalman filter track model in a global $\chi^2$ fit,
the $\chi^2$ would contain explicit contributions for the difference
in the state vectors at neighbouring nodes,
\[
\chi^2_{k-1,k} \; = \; \left ( x_k - F_{k-1} x_{k-1} \right)^{T} \left( Q_{k-1} \right)^{-1} \left ( x_k - F_{k-1} x_{k-1} \right) .
\]
These contributions are equivalent to the terms that constrain
scattering angles in the conventional track model for a global track
fit. As they represent additional constraints to the $\chi^2$, they
must also appear in the matrix $V$ and the residual vector $r$
in~\rfe{firstdchisqda}. It is only because of the minimum $\chi^2$
condition for the track parameters that their contribution in the
derivatives to the alignment parameters vanishes.


\section{Vertex constraints \label{sec:vertexconstraints}}

The expressions in~\rfe{propagation} can also be used to include
vertex or mass constraints in an alignment procedure. First, we
propagate the track parameters to the estimated position of the
vertex. We label the track parameters at that position with $x_0^n$
and its covariance by $C_0^{n}$. The correlations between these track
parameters and those at the position of each measurement can be
computed with the procedure outlined in the previous section. 

For clarity we now drop the superscript $n$ and replace it with a
superscript $(i)$ that labels the track in the vertex: The state of
track $i$ at the vertex is $x_0^{(i)}$ with covariance $C_0^{(i)}$. As
a result of the vertex fit (which we can implement as the
Billoir-Fr\"uhwirth-Regler
algorithm~\cite{Billoir:1985nq,Fruhwirth:1987fm}) we obtain the new
`constrained' track parameters $\tilde{x}_0^{(i)}$ with covariance
$\tilde{C}_0^{(i)}$.  The change in the track parameters can be
propagated to the track states at each measurement
using~\rfe{propagation}, which gives for the state vector at node $k$
\begin{equation}
  \tilde{x}_k^{(i)} \; = \; x_k^{(i)} \: + \: C_{k,0}^{(i)}
  \left(C_{0}^{(i)}\right)^{-1} ( \tilde{x}_0^{(i)} - x_0^{(i)} )
\end{equation}
and for the covariance 
\begin{equation}
  \tilde{C}_{k,\ell}^{(i)} \; = \; C_{k,\ell}^{(i)} + C_{k,0}^{(i)} 
   \left(C_0^{(i)}\right)^{-1} (\tilde{C}_0^{(i)} - C_0^{(i)} ) \left(C_0^{(i)}\right)^{-1} C_{0,\ell}^{(i)} .
\end{equation}
The constrained residuals for track $j$ then become
\begin{equation}
  \tilde{r}_k^{(i)} \; = \; m_k^{(i)} - h( \tilde{x}_k^{(i)} ) 
\end{equation}
and the covariance matrix $R$ in~\rfe{kalmanfullR} can be computed
using the new track state covariance $\tilde{C}_{k,\ell}^{(i)}$.

The vertex fit also gives us the covariance $\tilde{C}^{(i,j)}$
between any two tracks $i$ and $j$ in the vertex. This allows to
compute the correlation between any two states in any two tracks as
follows
\begin{equation}
  \tilde{C}_{k,\ell}^{(i,j)} \; = \; 
  C_{k,0}^{(i)} 
  \left( C_0^{(i)} \right)^{-1} \tilde{C}_0^{(i,j)}
  \left(C_0^{(j)}\right)^{-1} C_{0,\ell}^{(j)}
\end{equation}
Inserting this into the multi-track equivalent of~\rfe{kalmanfullR}
gives the full correlation matrix for the residuals on \emph{all}
tracks. If the number of tracks in the vertex is large, the
computation of the global covariance matrix for all states on all
tracks is rather CPU time consuming. Therefore, in practical
applications it makes sense to compute the correlation only for a
subset of hits close to the vertex.

This completes the ingredients for including vertex constraints in the
calculation of the alignment derivatives. Eventual mass constraints or
other kinematic constraints are included implicitly if they are
applied during the vertex fit.


\section{Application to the alignment of the LHCb tracking system \label{sec:application}}

The LHCb tracking system consists of a silicon vertex detector (VELO)
and a spectrometer~\cite{LHCbDetector}. For the track based alignment
of this system a closed-form alignment algorithm has been implemented in
the LHCb software framework. This algorithm, which uses the standard
LHCb Kalman filter track fit, will be described in detail in a future
publication~\cite{Amoraal:2008}. Here we briefly illustrate the effect
of correlations between residuals and the applications of vertex
constraints, using the alignment of the VELO system as an
example.\footnote{An alternative algorithm for the VELO alignment is
  described in~\cite{Viret:2008jq}.  It uses a standalone global track
  fit with a straight line track model, suitable only in the
  field-free region of the detector.}

The VELO system consists of 21 layers of double sided silicon
detectors with radial strips on one side and concentric circular
strips on the other. Each layer consist of two half circular disks
called \emph{modules}.  The modules are mounted onto two separate
support structures, the \emph{left} and \emph{right} VELO halves. The
two halves can be moved independently in the direction perpendicular
to the beam ($z$) axis in order to ensure the safety of the detectors
during beam injection.  The alignment of the VELO system is of crucial
importance to the physics performance of the LHCb experiment.

For the analysis described here we have simulated deformations in the
VELO detector in such a way that it bows along the $z$-axis: we
introduced a bias in the $x$ and $y$ position of each module that was
approximately proportional to $(z-z_0)^2$, where \mbox{$z-z_0$} was
the $z$ position of the module relative to the middle of the VELO. The
reason to choose this particular misalignment is that it corresponds
to a correlated movement of detector elements: Such deformations ---
sometimes called `weak modes' --- are inherently difficult to correct
for with an alignment method that ignores the correlations between
residuals in the track fit.

Tracks from a sample of simulated minimum bias interactions were
reconstructed using a `cheated' pattern recognition, assigning VELO
hits to a track based on the Monte Carlo truth. We required at least 8
hits per track.  The tracks were fitted with the standard LHCb track
fit, taking scattering corrections into account as process noise.
Tracks were accepted for alignment if their $\chi^2$ per degree of
freedom was less than 20. In a perfectly aligned detector this cut
only excludes a tiny fraction of reconstructed tracks, namely those
with a kink due to a hadronic interaction. Primary vertices were
reconstructed using the standard LHCb primary vertex finder.

To validate the implementation of the algorithm we have performed two
tests. First, we have checked the calculation of the residual
covariance matrix $R$ by comparing it to a numerical computation. A
single track was refitted after changing one measurement coordinate
$m_i$ by a numerically small value. The $i$-th row of $R$ follows from
(for $j \neq i$)
\[
R_{ij} \; = \; - \frac{\delta r_j^{(i)}}{\delta r_i^{(i)}} \: R_{ii},
\]
where $\delta r_j^{(i)}$ is the change in residual $r_j$. (The
computation of the diagonal element $R_{ii}$ is part of the standard
Kalman fit procedure.) This test has shown that the numerical
uncertainty in the correlations coefficients of $R$ is typically of
order $10^{-4}$, which is good enough for the purpose of detector
alignment.

Second, we have analyzed the eigenvalue spectrum of the second
derivative matrix~\rfe{seconddchisqda}. Without an external reference
system the global translations and rotations of a tracking system are
unconstrained in the alignment procedure. Such unconstrained degrees
of freedom lead to vanishing eigenvalues in the derivative matrix and,
if left untreated, result in a poorly converging alignment. (See
\eg{}~\cite{Bruckman:2005}.)  Unconstrained degrees of freedom can be
removed with Lagrange constraints or by omitting the corresponding
eigenvector from the solution to the linear system
in~\rfe{globalequation}.  However, to test the implementation of the
calculations in the global alignment algorithm, the identification of
the vanishing eigenvalues is a powerful tool: If the zero eigenvalues
corresponding to the global movements are observed, we can be
confident that the computation of both the matrix $R$ and the
alignment derivatives $\partial r / \partial\apar$ is correct (or at
least consistently wrong).

\begin{figure}
  \centerline{\includegraphics[width=0.48\textwidth]{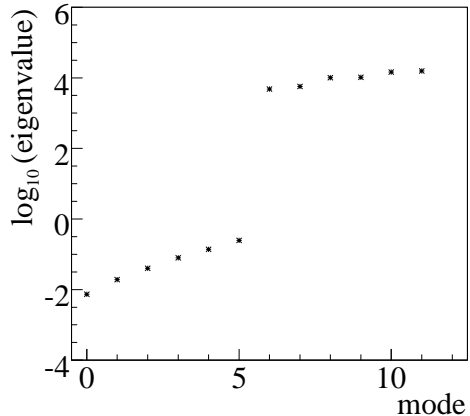}}
  \caption{Logarithm of the eigenvalue versus eigenvalue index
    (`mode') in the alignment of the position and rotation of two
    halves of the LHCb VELO.}
  \label{fig:evhalves}
\end{figure}

Figure~\rff{evhalves} shows the eigenvalues for the alignment of the
position and rotation of the two VELO halves. (The eigenvalues are
plotted versus an arbitrary index that increases with the size of the
eigenvalue.) The total number of alignment parameters is 12. To define
the scale of the eigenvalues the derivative matrix was rescaled
following the recipe in~\cite{Bocci:2007zzb}: the numerical value of the
eigenvalue is roughly equal to the number of hits contributing to the
corresponding linear combination of alignment parameters. As can be
seen in the figure the eigenvalue distribution splits in two: The six
smaller eigenvalues correspond to the global rotation and translation,
whereas the six larger eigenvalues correspond to the relative
alignment of the two detector halves. Note that if correlations
between residuals are ignored, the linear
equations~\rfe{globalequation} split in independent parts for the two
aligned objects and all eigenvalues are of about the same size.

One may wonder why the eigenvalues corresponding to the global
movements are not `numerically' zero, in contrast with the analysis
reported in~\cite{Bruckman:2005}. The reason for this is a feature of
the Kalman filter: In the Kalman filter the state vector is seeded
with a finite variance even before a single measurement is processed.
The variance must be large enough to have negligible weight in the
variance of the state vector after all measurements are processed, but
it must be small enough to make the computation of the filter gain
matrix numerically stable. The finite value of the seed variance
essentially fixes the track in space. We have observed that the value
of the small eigenvalues is indeed sensitive to the variance of the
seed. For practical purposes the bias from the Kalman filter seed is
not important.

\begin{figure}
  \centerline{\includegraphics[width=0.48\textwidth]{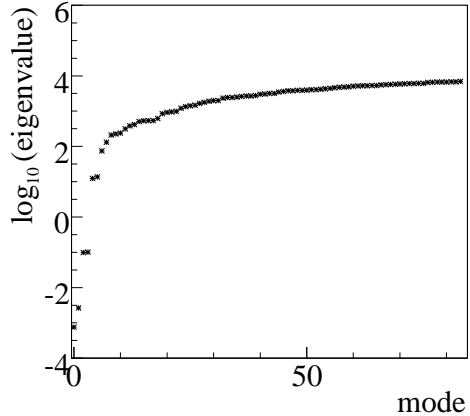}}
  \caption{Logarithm of the eigenvalue versus eigenvalue index
    (`mode') in the alignment of the position of modules in the LHCb
    VELO.}
  \label{fig:evmodules}
\end{figure}

To test the alignment procedure for the misalignment scenario
presented above we aligned the position of each module in $x$ and $y$,
corresponding to a total of 84 alignment parameters.  We omitted the
$z$ translation and rotations to simplify the analysis.  The
eigenvalue distribution, shown in figure~\rff{evmodules}, reveals 4
unconstrained degrees of freedom. These correspond to the global
translation in $x$ and $y$ and --- originating from the planar
geometry of the detector --- shearings in the $xz$ and $yz$ plane. We
constrain these degrees of freedom with Lagrange constraints.

We report here two figures of merit that we use to judge the
convergence of the alignment procedure, namely the number of selected
tracks and the average $\chi^2$ of selected tracks, both as a function
of the alignment iteration. The results are shown in
figure~\rff{convergence} for 3 different scenarios: First, we entirely
ignore correlations between residuals, which means that off-diagonal
elements in the matrix $R$ in equation~\rfe{residualcovariance} are
assumed zero. Second, we compute these correlations with the recipe
outlined in section~\ref{sec:kalmancorrelation}.  Finally, we also
include vertex constraints with the expressions given in
section~\ref{sec:vertexconstraints}. As can be seen in the figure the
scenario with correlations converges faster than the scenario without.
Furthermore, in the scenario without correlations less tracks survive
the $\chi^2$ cut even after 5 iterations.

\begin{figure}
  \includegraphics[width=0.48\textwidth]{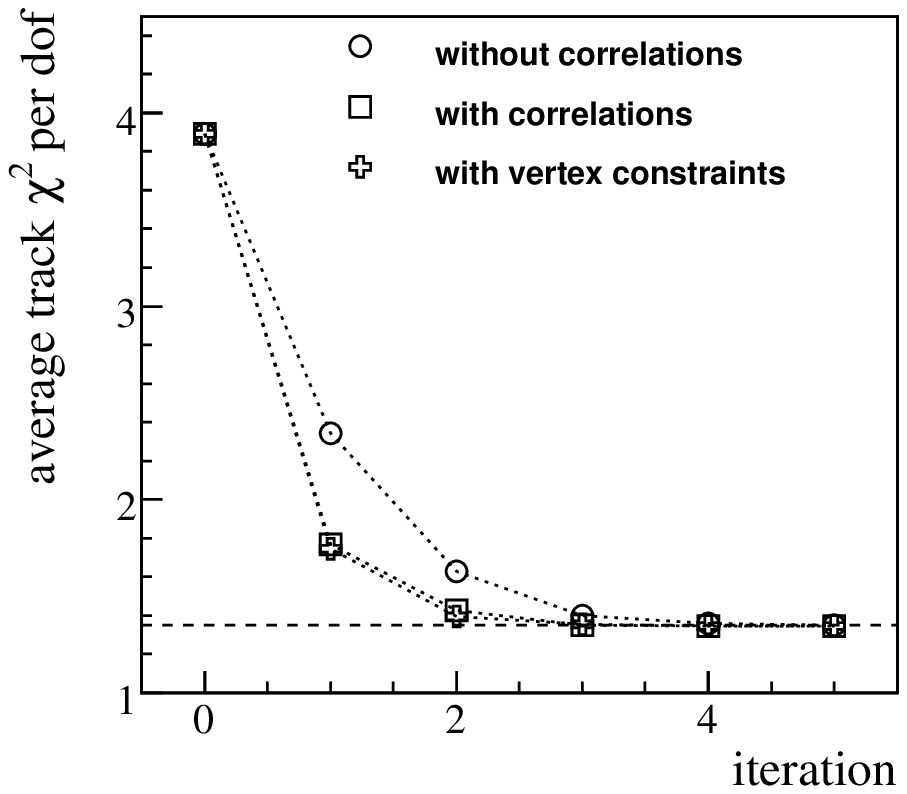}
  \includegraphics[width=0.48\textwidth]{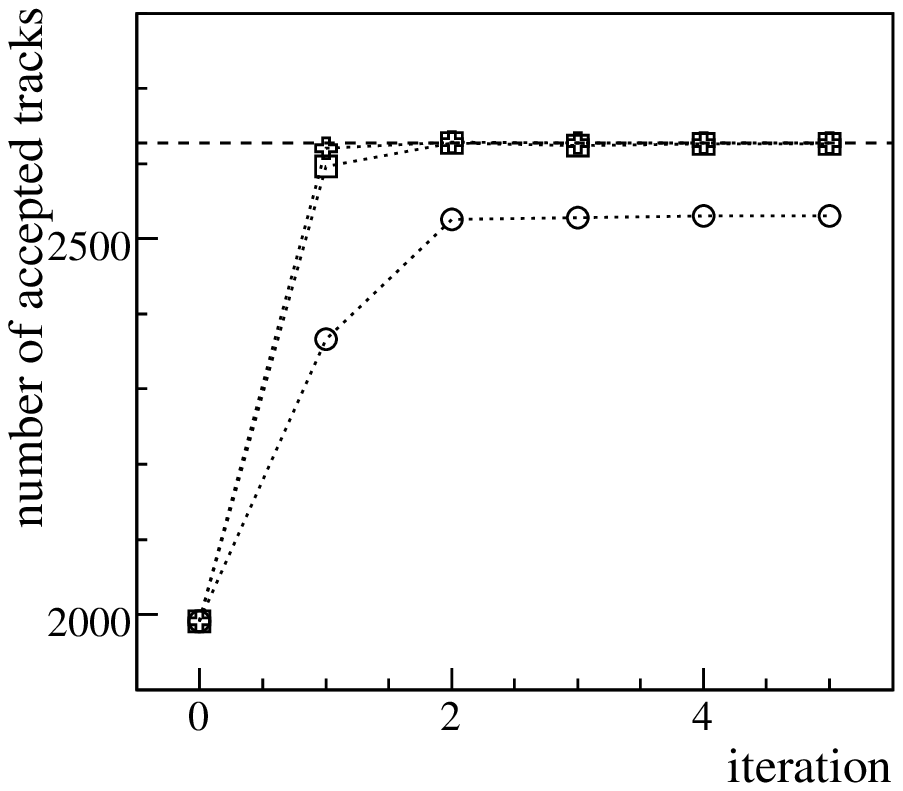}
  \caption{Number of selected tracks (right) and average $\chi^2$ per track (left) as a function of the number of alignment iterations for 3 alignment scenarios, namely ignoring correlations between residuals, not ignoring those correlations and including vertex constraints. The dashed line represents the result for a perfectly aligned detector.}
  \label{fig:convergence}
\end{figure}
 
The difference in convergence behaviour is mostly because there are
two kinds of tracks. Though most tracks pass only through a single
VELO half, there is a small fraction that passes through small regions
in which detectors from both halves overlap. When correlations between
hit residuals are taken into account, tracks that pass through a
single half do not carry any weight in determining the relative
positions of the two halves, because the contribution to the $\chi^2$
is invariant to the global position of the detector half.  Therefore,
the relative position of the two halves is fully sensitive to the
tracks that pass through both halves. On the contrary, if correlations
are ignored, every track fixes the position of any detector element in
space. As a result the overlap tracks get a much smaller weight in
determining the relative position and convergence becomes poor.

This problem can be partially overcome by explicitly enhancing the
fraction of overlap tracks in the sample, \eg{} by down-sampling the
tracks that do not pass through the overlap regions.  Such a strategy
is applied in the alignment of the Babar vertex
detector~\cite{Brown:2008cc}. An important advantage of the
closed-form algorithm is that it is not necessary to remove tracks
with a small weight in the alignment as the algorithm inherently
weights the information contained in the residuals correctly.


\section{Conclusions}

In this paper we have presented how the most popular track fitting
method, the Kalman filter, can be used in a closed-form alignment
procedure for tracking detectors.  Our contribution is summarized in
expression~\rfe{recursivecorrelation} which shows how the correlations
between state vectors can be computed recursively by using the
smoother gain matrix. We have also shown how vertex constraints can be
included without refitting the tracks.  Using an implementation of
this formalism in the LHCb software framework we have illustrated for
a simple misalignment scenario of the LHCb vertex detector the
importance of correlations between residuals in the track fit. A more
detailed analysis of the performance of the alignment algorithm to the
LHCb tracking system will be reported in due
course~\cite{Amoraal:2008}.

\section*{Acknowledgments}

The author would like to thank G.~Raven for posing the question
concerning correlations between residuals in the Kalman filter and for
his patient proof reading of this manuscript. The alignment software
that was used for the analysis in section~\ref{sec:application} was
developed in close collaboration with J.~Amoraal, A.~Hicheur,
M.~Needham and L.~Nicolas and G.~Raven,

\end{document}